\documentclass[aps,twocolumn,floats,prl,nofootinbib]{revtex4-1}

\usepackage[dvips]{graphicx} %
\usepackage{graphicx,amsmath,amsfonts,amssymb,slashed}
\usepackage{bbold,wasysym}
\usepackage{graphicx}
\usepackage{txfonts}

\usepackage[usenames,dvipsnames]{xcolor} 

\usepackage{soul}

\definecolor{RedWine}{rgb}{0.743,0,0}
\definecolor{RoyalBlue}{rgb}{0.25,.41,.88}

\setstcolor{Blue}

\newcommand{\kD}{k_{\rm D}}
\newcommand{\kDmu}{k_{\rm D, \mu}}
\newcommand{\kDmuy}{k_{\rm D, \mu y}}
\newcommand{\kDy}{k_{\rm D, y}}

\newcommand{\Mpc}{\ensuremath{\mathrm{Mpc}}}

\newcommand{\fnl}{\ensuremath {f_{\mathrm{nl}}}}

\def\VEV#1{\left\langle #1 \right\rangle}

\begin{document}

\title{Probing the scale dependence of non-Gaussianity with
     spectral distortions of the cosmic microwave background}

\author{Razieh Emami$^{1,2}$, Emanuela Dimastrogiovanni$^3$, Jens
     Chluba$^4$, and Marc Kamionkowski$^2$}
\affiliation{$^1$School of Physics, Institute for Research in
     Fundamental Sciences (IPM) P. O. Box 19395-5531, Tehran,
     Iran \\
     $^2$Department of Physics and Astronomy, Johns Hopkins
     University, 3400 N.\ Charles St., Baltimore, MD 21218, USA \\
     $^3$Department of Physics and School of Earth and Space Exploration Arizona State University, Tempe, AZ 85827, USA\\
     $^4$Institute for Astronomy, K30, University of Cambridge,
     Madingley Road, Cambridge CB3 0HA, United Kingdom}

\begin{abstract}
Many inflation models predict that primordial density
perturbations have a nonzero three-point correlation function, or
bispectrum in Fourier space.  Of the several possibilities for
this bispectrum, the most commmon is the local-model bispectrum,
which can be described as a spatial modulation of the small-scale
(large-wavenumber) power spectrum by long-wavelength density
fluctuations.  While the local model
predicts this spatial modulation to be scale-independent, many
variants have some scale-dependence.  Here we note that this scale
dependence can be probed with measurements of frequency-spectrum
distortions in the cosmic microwave background (CMB), in
particular highlighting Compton-$y$ distortions.
Dissipation of primordial perturbations with wavenumbers
$50\,\Mpc^{-1} \lesssim  k \lesssim 10^4\,\Mpc^{-1}$ give rise to
chemical-potential ($\mu$) distortions, while those with
wavenumbers $1\,\Mpc^{-1} \lesssim k \lesssim 50\,\Mpc^{-1}$
give rise to Compton-$y$ distortions.  With local-model
non-Gaussianity, the distortions induced by this dissipation can
be distinguished from those due to other sources via their
cross-correlation with the CMB temperature $T$.  We show that
the relative strengths of the $\mu T$ and $yT$ correlations thus
probe the scale-dependence of non-Gaussianity and
estimate the magnitude of possible signals relative to
sensitivities of future experiments.  We discuss the
complementarity of these measurements with other probes of
squeezed-limit non-Gaussianity.
\end{abstract}

\maketitle

The prevailing paradigm for the origin of the primordial density
perturbations inferred from fluctuations in the cosmic microwave
background (CMB) and from large-scale galaxy surveys is that
they arose as quantum fluctuations during a quasi-de-Sitter
phase (``inflation'') of expansion in the early Universe
\cite{inflation}.  Inflation
generally predicts a spectrum of primordial
perturbations that is nearly, but not precisely,
scale-invariant, consistent with current measurements.  It also
generically predicts that these perturbations should be very
close to, but not precisely, Gaussian (i.e., have only small
connected correlations beyond the two-point correlation
function).  The amplitude and detailed form of this
non-Gaussianity varies considerably between models and is
usually specified in terms of a three-point function, or
``bispectrum'' in Fourier space \cite{NGReview}.

The {\it local model} bispectrum, which peaks in the squeezed
limit $k_2\simeq k_3 \gg k_1$ (where $k_1\leq k_2 \leq k_3$ are
the three Fourier modes being correlated) appears in a number of
simple inflation models \cite{localmodel,Verde:1999ij} and has
thus become a standard
workhorse.  Local-model non-Gaussianity has been sought in the
CMB \cite{Verde:1999ij,Komatsu:2008hk,Ade:2013ydc,Ade:2015ava}, in
large-scale structure \cite{Verde:2001sf}, and through the
scale-dependent biasing
\cite{Dalal:2007cu,Slosar:2008hx,Matarrese:2008nc,Schmidt:2010gw}
it induces in galaxy clustering at large scales.

It has also recently been shown \cite{Pajer:2012vz,Ganc:2012ae}
that local-model non-Gaussianity
will induce spatial correlations between the CMB
temperature fluctuation and chemical-potential ($\mu$)
distortions as a function of position on the sky.  These $\mu$
distortions arise from dissipation of primordial perturbations
with wavenumbers $50\,\Mpc^{-1} \lesssim  k \lesssim
10^4\,\Mpc^{-1}$ at redshifts $z\simeq 5\times10^4 - 2\times
10^6$ \cite{Hu:1994bz,Chluba:2012gq,Chluba:2012we}.  In the
local model, the amplitude of these small-scale
perturbations is modulated by the long-wavelength curvature
perturbation that also gives rise to large-angle temperature
fluctuations, thus inducing a $\mu T$ correlation
\cite{Pajer:2012vz,Ganc:2012ae}.  These $\mu T$
correlations therefore probe local-model non-Gaussianity at wavenumbers
$50\,\Mpc^{-1} \lesssim  k \lesssim 10^4\,\Mpc^{-1}$ far smaller
than those accessible with CMB temperature fluctuations,
galaxy surveys, or even future 21cm observations
\cite{Loeb:2003ya}.

Still, the local-model bispectrum is just one of an infinitude
of different types of bispectra.  Large
classes of bispectra have arisen from different ideas for
inflation \cite{larger,curvaton,Dvali:1998pa,topdefects},
and no shortage of phenomenological parametrizations have been
considered.

In this paper we point out that the correlation of CMB Compton-$y$
distortions with the large-angle temperature fluctuation can be
used, in tandem with the $\mu$T correlation, to probe different
types of non-Gaussianity.  Dissipation of primordial
perturbations with wavenumbers\footnote{While
modes with $k\lesssim 1\,\Mpc^{-1}$ also damp and create a
$y$-distortion, after recombination the effective heating rate
for these perturbations is much lower than at smaller scales
\cite{Chluba:2012gq,Chluba:2012we}.}\ $1\,\Mpc^{-1} \lesssim
k \lesssim 
50\,\Mpc^{-1}$ gives rise to Compton-$y$ distortions to the CMB
\cite{Chluba:2012gq}.
However, these spectral  distortions have been largely
disregarded because they are not easily distinguished from
larger Compton-$y$ signals from the intergalactic medium at low
redshifts.  Non-Gaussianity may, however, induce a $yT$
correlation that could allow this dissipation-induced $y$
distortion to be isolated from late-time effects
(e.g., \cite{Cen:1998hc, Zhang:2004fh}) by means of its angular
dependence. But even without being able to distinguish different
contributions, one can still derive conservative upper limits
using $yT$ correlations.  Since
$y$ and $\mu$ distortions probe different wavenumbers, the
relative strength of the $yT$ and $\mu T$ distortions can be
used to probe the scale dependence of  primordial
non-Gaussianity, extending previous considerations of just the
$\mu T$ correlation \citep{Biagetti:2013sr}.

Below, we first present a simple calculation, based upon the
configuration-space description (rather than the Fourier-space
description in prior work \cite{Pajer:2012vz,Ganc:2012ae}) of
local-model non-Gaussianity and the $\mu T$ correlation.  This
calculation illustrates clearly that the $\mu T$ correlation
arises from large-scale modulation of small-scale modes.  The
generalization to $yT$ correlations is thus clear.  We then
parametrize the type of non-Gaussianity probed by the
combination of $\mu T$ and $yT$ correlations and estimate the
sensitivity of future experiments.  We close by discussing the
complementarity of these measurements with other probes of
squeezed-limit non-Gaussianity.

To begin, the average $\mu$ and $y$ distortions induced, with Gaussian
initial conditions, by dissipation of primordial perturbations
are given by
\begin{subequations}
\label{eqn:muysource}
\begin{align}
   \VEV{\mu} &\approx \int {\rm d}\log k \, \Delta_{\cal R}^2(k)  W_\mu(k),
\label{eqn:musource}
\\
   \VEV{y} &\approx  \int {\rm d}\log k \, \Delta_{\cal R}^2(k) W_y(k),
\label{eqn:ysource}
\end{align}
\end{subequations}
where $\Delta_{\cal R}^2(k)$ is the primordial curvature power
spectrum, and $W_\mu(k)$ and $W_y(k)$ define $k$-space window
functions to account for the acoustic-damping and thermalization
physics. Although more accurate expressions for the $k$-space
window functions have been discussed
\cite{Chluba:2012we,Chluba:2013dna,Chluba:2015}, here we will
use the simple approximations,
\begin{subequations}
\begin{align}
W_\mu(k)&\approx 2.3\left[{\rm e}^{-2k^2/\kDmu^2}-{\rm e}^{-2k^2/\kDmuy^2}\right] ,
\\
W_y(k) &\approx 0.4 \left[{\rm e}^{-2k^2/\kDmuy^2}-{\rm e}^{-2k^2/\kDy^2}\right] ,
\end{align}
\end{subequations}
where $\kD(z)\simeq 4.1\times 10^{-6} \, (1+z)^{3/2}\,\Mpc^{-1}$ is the diffusion damping scale at redshift $z$.
This quantity is evaluated at the initial and final redshifts at which $\mu$ and $y$ distortions are produced, assuming that the transition between $\mu$ and $y$ happens abruptly at $z\simeq 5\times10^4$ \cite{Hu:1992dc,Chluba:2011hw}. Thus, 
$\kDmu=\kD(2\times 10^6)\simeq 1.1\times10^4\, \Mpc^{-1}$,
$\kDmuy =\kD(5\times 10^4) \simeq 46\, \Mpc^{-1}$, and
$\kDy=\kD(1090) \simeq 0.15\, \Mpc^{-1}$.
Evaluating the expressions in Eq.~\eqref{eqn:muysource} using $\Delta_{\cal R}^2(k) \simeq 2.4\times
10^{-9} \, (k/0.002\,\Mpc^{-1})^{n_s-1}$, with $n_s\simeq0.96$
\cite{Komatsu:2010fb}, yields $\VEV{\mu} \simeq 1.9\times 10^{-8}$
and $\VEV{y}\simeq 4.2\times 10^{-9}$, in good agreement with
more detailed computations \cite{Chluba:2012gq}.

Since the angle subtended by a causal region at the surface of
last scatter is $\sim1^\circ$, we see in the CMB $\sim40,000$
causally disconnected Universes.  If primordial perturbations
are Gaussian, then the amplitude $\Delta_{\cal R}^2(k_{\rm
small})$ of primordial perturbations on the small scales $k_{\rm
small}$ that induce spectral distortions will be the same
everywhere.  If, however, there is local-model non-Gaussianity,
then the power spectrum $\Delta_{\cal R}^2(k_s,\vec x)$ for
small-scale modes $k_s$ will differ from one causal patch
centered at position $\vec x$ to another.  The fluctuation will
moreover be correlated with the long-wavelength curvature
fluctuation ${\cal R}(\vec x)$.

This can be understood simply from the configuration-space
description of local-model non-Gaussianity.  The local-model
curvature perturbation at position $\vec x$ is written,
\begin{equation}
     {\cal R}(\vec x) = r(\vec x )+ \frac{3}{5} \fnl r^2(\vec
     x),
\end{equation}
in terms of a Gaussian random variable $r(\vec x)$.  We then
write $ {\cal R}(\vec x)= {\cal R}_L(\vec x)+  {\cal R}_s(\vec x)$, where $ {\cal R}_L(\vec x)$
is the part of $ {\cal R}(\vec x)$ that comes from long-wavelength
Fourier modes and $ {\cal R}_s(\vec x)$ that from short-wavelength
Fourier modes, and similarly write $r(\vec x) = r_L(\vec
x)+r_s(\vec x)$.  By writing
\begin{equation}
      {\cal R}_L +  {\cal R}_S = r_L + r_s + \frac35 \fnl
      \left[r_L^2 +2 r_L r_s +
     r_s^2 \right],
\end{equation}
we infer that the small-scale curvature fluctuation in the
presence of some fixed long-wavelength curvature fluctuation
$ {\cal R}_L(\vec x)$ is, to linear order in $\fnl$,
\begin{equation}
      {\cal R}_s(\vec x) = r_s(\vec x) \left[ 1 +  \frac{6}{5} \fnl
      {\cal R}_L(\vec x) \right].
\end{equation}
Thus, the fractional change in small-scale power in a region of
a fixed long-wavelength curvature fluctuation is
\begin{equation}
     \frac{\delta \VEV{{\cal R}^2}}{ \VEV{{\cal R}^2}} \approx
     \frac{12}{5} \fnl R_L(\vec x).
\end{equation}
We thus infer that, with local-model non-Gaussianity, the
(fractional) chemical-potential fluctuation in a given region of
the sky is given simply by the long-wavelength curvature
perturbation in that region at the surface of last scatter.

The same is true for the large-angle temperature
fluctuation---it is determined primarily by the curvature
fluctuation at the surface of last scatter and has magnitude
$\Delta T/T \approx {\cal R}/5$.  Therefore, for
multipole moments $\ell \lesssim 100$ that probe causally
disconnected regions at the surface of last scatter, the
cross-correlation between the (fractional) chemical-potential
fluctuation $\Delta \mu/\mu \approx \delta \VEV{{\cal
R}^2}/\VEV{{\cal R}^2}$ and the temperature fluctuation $\Delta
T/T$ has a power spectrum,
\begin{equation}
     C_\ell^{\mu T} \approx 12\, \fnl C_\ell^{TT}.
\end{equation}
This easily obtained result agrees with
Ref.~\cite{Pajer:2012vz}, noting that their $C_\ell^{\mu T}$ is
for the $\mu$ fluctuation, rather than the fractional $\mu$
fluctuation.  The $\mu$ autocorrelation caused by
non-Gaussianity is $C_\ell^{\mu\mu} \approx
144\, \fnl^2 C_\ell^{TT}$.  Here we assumed that the trispectrum
contributions are negligible.  As our derivation clarifies, the
$\mu$T correlation arises from the squeezed limit of the
bispectrum, the part of the bispectrum that modulates
small-scale power.  For completeness, we include the
large-angle temperature power spectrum,
\begin{equation}
     C_\ell^{TT} = \frac{2\pi}{25} \frac{\Delta_{\cal
     R}^2}{\ell(\ell+1)} \simeq \frac{6.0\times
     10^{-10}}{\ell(\ell+1)}.
\end{equation}
If the non-Gaussianity is scale-invariant, the $yT$ and $yy$
correlations are the same as the $\mu T$ and $\mu\mu$
correlations (all in terms of $\Delta y/y$ and $\Delta \mu/\mu$)
\begin{equation}
     C_\ell^{y T} \approx 12\, \fnl C_\ell^{TT}\,,\quad\quad
     C_\ell^{yy} \approx 144\, \fnl^2 C_\ell^{TT}.
\end{equation}

If, however, the non-Gaussianity is scale-dependent (see, e.g.,
Refs.~\cite{LoVerde:2007ri,Byrnes:2010ft,Byrnes:2009pe,Bramante:2011zr,Riotto:2010nh,Becker:2012je,Sefusatti:2009xu}),
then the value of $\fnl$ that describes correlations between the long-wavelength
($k_L\lesssim0.01\,\Mpc^{-1}$) modes responsible for large-angle
CMB temperature fluctuations and short-wavelength modes
($50\,\Mpc^{-1} \lesssim k \lesssim 10^4\,\Mpc^{-1}$)
responsible for the $\mu$ distortion, may differ from the value
of $\fnl$ that
describes correlations between long-wavelength modes and
$1\,\Mpc^{-1} \lesssim k \lesssim 50\,\Mpc^{-1}$ modes
responsible for the dissipation-induced $y$ distortions.  

We therefore parametrize the scale-dependent non-Gaussianity in
terms of a scale-dependent non-Gaussianity parameter $\fnl(k_s)$
defined by the squeezed-limit ($k_{s}\equiv k_{2}\simeq
k_{3}\gg k_{1}\equiv k_L $) curvature bispectrum,
\begin{equation}
     B_{\cal R}(k_1,k_2,k_3) \simeq \frac{12}{5} \fnl(k_s)
     P_{\cal{R}}(k_{s})P_{\cal{R}}(k_{L}),
\label{eqn:Blimit}
\end{equation}
defined from
\begin{equation}
     \VEV{\mathcal{R}_{\vec{k}_{1}} \mathcal{R}_{\vec{k}_{2}}
     \mathcal{R}_{\vec{k}_{3}}} \equiv
     (2\pi)^{3}\delta^{(3)}(\vec{k}_{1}+\vec{k}_{2}+\vec{k}_{3})
     B_{\cal R}(k_1,k_2,k_3)
\end{equation}
for $k_L \lesssim 0.01\,\Mpc^{-1}$. We then define two
parameters, $\fnl^y \equiv \fnl(k_s\simeq
7\,\Mpc^{-1})$ and $\fnl^\mu \equiv \fnl(k_s\simeq
740\,\Mpc^{-1})$ to parametrize the non-Gaussianity on $\mu$
and $y$ scales, respectively.  Here, the \fnl\ parameter
is evaluated roughly at the log-midpoints of the $y$- and
$\mu$-distortion $k$ intervals, which defines the $y$- and
$\mu$-distortion pivot wavenumbers $k_y\simeq 7\,\Mpc^{-1}$ and
$k_\mu\simeq 740\,\Mpc^{-1}$, respectively.  Although the
scale-dependence of $\fnl$ is often modeled in the literature as
a power law in wavenumber $k$, this parametrization does not
accommodate the possibility, which arises in curvaton and
multi-field models
\cite{Bramante:2011zr,Enqvist:2005pg,Byrnes:2011gh,Meyers:2013gua},
that $\fnl$ may change sign with scale. The above
parametrization is therefore more general.
We note that the $k_1\equiv k_L \ll k_2 \simeq k_3 \equiv k_s$
bispectrum in Eq.~(\ref{eqn:Blimit}) is the squeezed limit 
of the common bispectrum parametrization (e.g.,
\cite{Becker:2010hx}), $B_{\cal R}(k_1,k_2,k_3) = 
(6/5) \left[\fnl(k_1) P(k_2)P(k_3) + {\rm perms} \right]$, of
scale-dependent non-Gaussianity.  Our results are thus easily
compared with many other results on scale-dependent
non-Gaussianity.

We estimate the detectability of the $\mu T$ signal following
Ref.~\cite{Pajer:2012vz} assuming a noise power spectrum
$C_\ell^{\mu,{\rm n}} \simeq w_\mu^{-1}
e^{\ell^2/\ell_{\rm max}^2}$ with $w_\mu^{-1/2}\approx\sqrt{4\pi}\,
(\mu_{\rm min}/\VEV{\mu})$ and $l_{\rm max}\simeq
100$. Note again that ours is a power
spectrum for $\Delta\mu/\mu$, as opposed to the power spectrum
for $\mu$ in Ref.~\cite{Pajer:2012vz}, where $\mu_{\rm
min}$ is the smallest detectable $\mu$ monopole, which is
estimated to be $\mu_{\rm min} \simeq 10^{-8}$ for PIXIE
\cite{Kogut:2011xw} and $\mu_{\rm min}\simeq 10^{-9}$ for PRISM
\cite{Andre:2013nfa}.  We then obtain the smallest detectable
(at $\sim 1\sigma$) $\mu T$ correlation to arise for
\begin{equation}
     \fnl^\mu\simeq 220\, \left( \frac{\mu_{\rm min}}{10^{-9}}
     \right)\left( \frac{\VEV{\mu}}{2\times10^{-8}}
     \right)^{-1}.
\end{equation}
Although the dissipation-induced average $y$ distortion is expected to
be $\simeq 5$ times 
smaller than the $\mu$ distortion (see estimate above), the
experimental sensitivity to $y$ is $\simeq 5-10$ times
better \cite{Kogut:2011xw,Chluba:2013pya}. Assuming a PRISM
detection limit of $y_{\rm min}\simeq 2\times 10^{-10}$, we thus
infer a comparable smallest non-Gaussianity
parameter,
\begin{equation}
     \fnl^y \simeq 220\, \left( \frac{y_{\rm min}}{2 \times 10^{-10}} \right)\left( \frac{\VEV{y}}{4\times10^{-9}} \right)^{-1},
\end{equation}
that will engender a detectable $yT$ correlation.
These estimates are obtained assuming a roughly scale-invariant
spectrum of primordial perturbations.  If for some reason the small-scale
power-spectrum amplitude is increased in the $y$ range ($1\,\Mpc^{-1} \lesssim  k \lesssim 50\,\Mpc^{-1}$) or $\mu$
range ($50\,\Mpc^{-1} \lesssim  k \lesssim 10^4\,\Mpc^{-1}$), the
smallest detectable $\fnl^y$ and $\fnl^\mu$, respectively, will
be decreased by a similar factor.  In this case, the
homogeneous values of $y$ and $\mu$ will also be increased
\cite{Chluba:2012we}. The detection limits obtained from the
$\mu$ and $y$ autocorrelations are typically a few times weaker,
since in contrast to the temperature cross-correlations noise
dominates the $\mu$ distortion measurements.

The $\mu$ and $y$ spectral distortions in any given causal patch
at the surface of last scatter come from the dissipation of
small-scale modes within that patch.  There is a finite
number $N\sim (k\lambda)^3$ of such modes in this patch, where
$k$ is the largest relevant wavenumber ($10^4\,\Mpc^{-1}$ and $50\,\Mpc^{-1}$
for $\mu$ and $y$ distortions, respectively), and
$\lambda\sim100\,\Mpc$ the the size of the causal patch at the
surface of last scatter.  There will therefore be Poisson
fluctuations of amplitude $\sim N^{-1/2}$ in the value of the
$\mu$ and $y$ distortions in any such patch.  The mean-square
fractional $\mu$ fluctuation in a patch of angular size $\theta$
will thus be,
\begin{equation}
     \left(\frac{\Delta \mu}{\mu} \right)_\theta^2 = \int \frac{
     \ell {\rm d}\ell}{2\pi} C_\ell |W_\ell|^2 \sim \left. \ell^2
     C_\ell\right|_{\ell\sim \theta^{-1}} \sim \left(\frac{H_0}{k_{\rm max}
     \theta}\right)^{3}\nonumber,
\end{equation}
where $W_\ell$ is the window function for a circle on the sky of
radius $\theta$, and the last (approximate) equality is the
Poisson-fluctuation amplitude.  We therefore infer fluctuations
$C_\ell^{yy}\sim 6\times 10^{-15}\,(\ell/100)$ and
$C_\ell^{\mu\mu}\sim 6\times 10^{-25}\,(\ell/100)$ well below
those from measurement noise.

\begin{figure}[htbp]
\centering
\includegraphics[width=0.48 \textwidth]{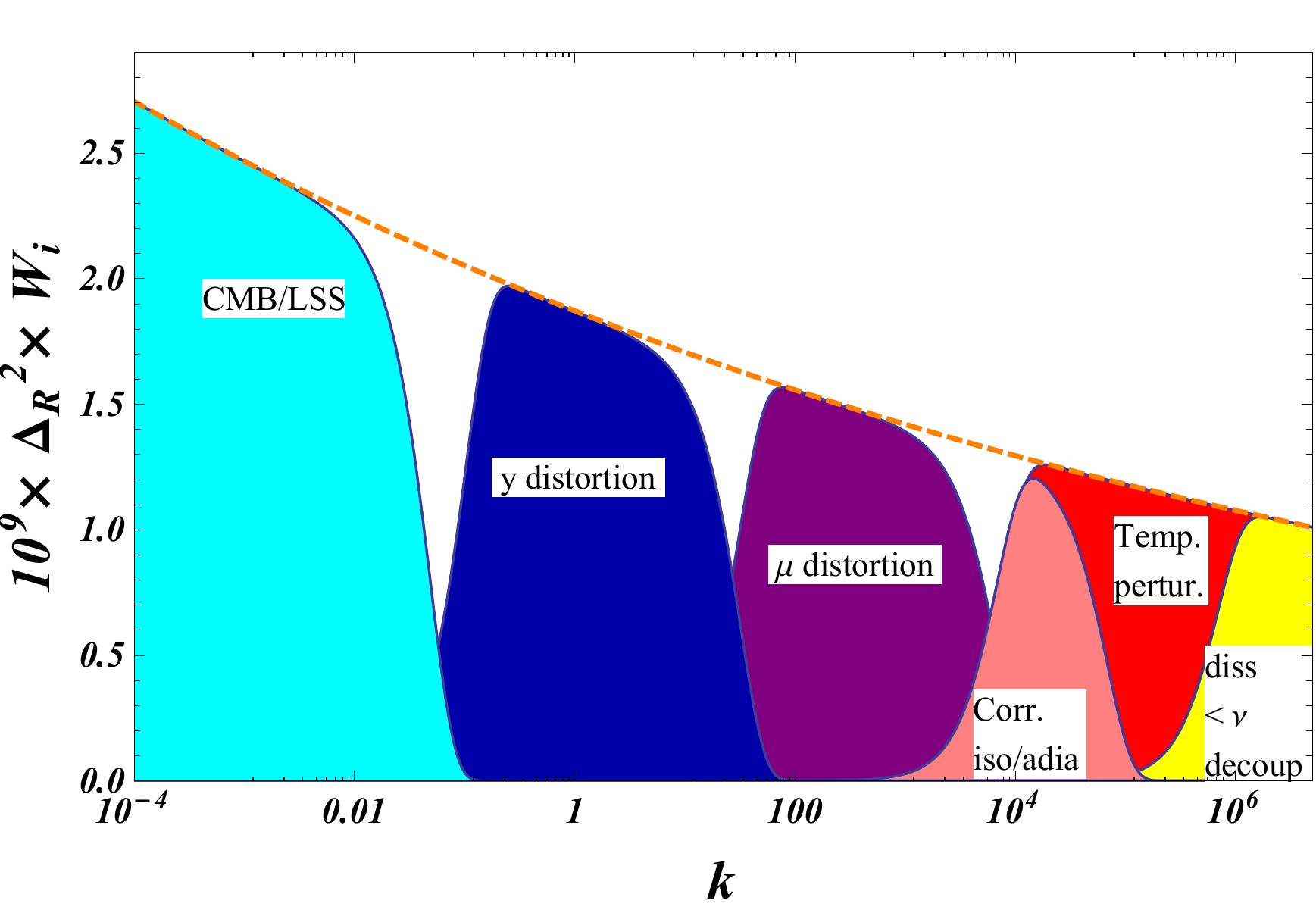}
\caption{The different $k$-space windows responsible for
     different observables. The CMB anisotropies are visible
     at $k\lesssim 0.1\, \Mpc^{-1}$, where the
     smale-scale cut-off is introduced by
     Silk damping. Compton-$y$-distortions are formed by the
     damping of modes with $1\,\Mpc^{-1}\lesssim k \lesssim
     50\,\Mpc^{-1}$, while $\mu$-distortions are created by
     modes with $50\,\Mpc^{-1}\lesssim k \lesssim
     10^4\,\Mpc^{-1}$. Additional large-scale temperature power
     can be created by modes at $10^4\,\Mpc^{-1}\lesssim
     k$ \protect\cite{Naruko:2015pva}. Those with
     $10^4\,\Mpc^{-1}\lesssim k\lesssim
     10^5\,\Mpc^{-1}$ can be constrained using the primordial
     helium abundance
     \protect\cite{Jeong:2014gna,Nakama:2014vla}, while modes
     with $10^5\,\Mpc^{-1}\lesssim
     k$ are erased before neutrino decoupling and thus would
     only lead additional large-scale temperature
     fluctuations.}
\label{fig:thefigure}
\end{figure}

Before the $yT$ correlations considered here can be used to
probe or constrain primordial non-Gaussianity, it will be
necessary to consider the cross-correlation of the Compton-$y$
distortion from intergalactic gas in the late Universe with the
contribution to the large-angle temperature fluctuation from the
integrated Sachs-Wolfe effect.  We anticipate that lensing
reconstruction can be used to separate out this late-time $yT$
correlation.  We also anticipate that this contribution can be
distinguished by a different $\ell$ dependence. 

An additional source of $yT$ correlations could arise from the
damping of primordial magnetic fields
\cite{Miyamoto:2013oua}. In spite of the large uncertainty in
the amplitude of primordial magnetic fields, the
scale-dependence is again generally expected to differ from the
one caused by non-Gaussianity. However, a more detailed study is
required.

The constraints to the squeezed-limit bispectrum from $\mu T$ and
$yT$ correlation considered here will be
complemented on the longer- and shorter-wavelength ends by other
measurements.  Searches for non-Gaussianity in CMB fluctuations
and in galaxy surveys probe wavenumbers $k\lesssim
0.1\,\Mpc^{-1}$.  There are other searches for the
scale-dependent bias that arises from non-Gaussianity
\cite{Dalal:2007cu,Slosar:2008hx,Matarrese:2008nc}.  These
probe squeezed-limit non-Gaussianity primarily on wavenumbers
$k\lesssim 1\,\Mpc^{-1}$ \cite{Schmidt:2010gw} (those that are
most important for determining the abundance of the galaxies
being correlated).  Dissipation of acoustic modes with
wavelengths $10^4\,\Mpc^{-1} \lesssim k\lesssim 10^5\,\Mpc^{-1}$
produce entropy in the primordial plasma after BBN
\cite{Jeong:2014gna,Nakama:2014vla}.  The long-wavelength
modulation induced by squeezed-limit non-Gaussianity on these
scales will then give rise to a small isocurvature fluctuation
correlated with the primordial adiabatic perturbation
\cite{Jeong:2014gna}. Finally, modes with $10^4\,\Mpc^{-1}
\lesssim k$ can give rise to additional large-scale temperature
fluctuations caused by non-Gaussianity
\cite{Naruko:2015pva}. Depending on the sign of the correlation,
this could also produce a lack of power on large scales (which
could also affect the values of cosmological parameters inferred
from these temperature fluctuations).

We find it interesting that there are thus now prospects to
probe the amplitude of squeezed-limit non-Gaussianity on a
continuum of distance scales from the largest ($\sim$Gpc)
accessible to those on scales nearly eight orders of magnitude
smaller, with CMB, large-scale structure, scale-dependent
biasing, $yT$ correlations, $\mu T$ correlations, and
small-scale entropy production.  This complement of
measurements will thus allow the determination of the functional
dependence of $\fnl(k_s)$, without necessarily assuming a
specific parametrization (e.g., power-law) for its
scale-dependence.

\begin{acknowledgments}
This work was supported at JHU by NSF Grant No.\ 0244990, NASA
NNX15AB18G, the John Templeton Foundation, and the Simons
Foundation, and at ASU by the Department of Energy.  JC was
supported by a Royal Society Research Fellowship, and RE
acknowledges the support of the New College Oxford - Johns
Hopkins Centre for Cosmological Studies.
\end{acknowledgments}

\end{document}